\documentclass[epjST,final]{svjour}
\usepackage{graphicx,amsfonts,amsmath,amssymb,xcolor,subfigure}
\usepackage[english]{babel}
\usepackage[numbers,sort&compress]{natbib}
\usepackage[latin1]{inputenc}
\usepackage{braket}
\usepackage{pifont}
\usepackage{xcolor}

\newcommand{\p}[2]{\frac{\partial\, #1}{\partial\, #2}\,}
\newcommand{\pt}[2]{\frac{\mathrm{d}\, #1}{\mathrm{d}\, #2}\,}

\newcommand{\eq}[1]{$\mathrm{Eq.}$~\eqref{#1}}

\newcommand{\secref}[1]{$\mathrm{Sec.}$~\ref{#1}}

\newcommand{\notiz}[1]{{\color{black} #1}}
\newcommand{\av}[1]{\left\langle #1 \right\rangle}

\begin{document}
\title{Eliminating Inertia in a Stochastic Model of a
  Micro-Swimmer with Constant Speed} \author{S. Milster \inst{1} \and
  J. N\"otel \inst{1} \and I. M. Sokolov \inst{1} \and L.
  Schimansky-Geier\inst{1,2}}

\institute{Department of Physics, Humboldt-University at Berlin,
  Newtonstr. 15, 12489 Berlin, Germany \and Department of Physics and
  Astronomy, Ohio University, Athens, Ohio 45701, USA}

\abstract{We \notiz{are concerned with the dynamical} description of the
  motion of a stochastic micro-swimmer with constant speed and
  fluctuating orientation in the long time limit by adiabatic
  elimination of the orientational variable.  Starting with the
  corresponding full set of Langevin equations, we eliminate the
  memory in the stochastic orientation and obtain a stochastic
  equation for the position alone in the overdamped limit. An
  equivalent procedure based on the Fokker-Planck equation is
  presented as well.}
\maketitle

\section{Introduction}
Recently, there has been an increasing interest in biology, physics
and chemistry in so-called active particles
\cite{niwa1994,vicsek1995,marchetti.c:2013,romanczuk.p:2012,hauser_spec}.
These units are equipped with some kind of propulsive mechanism which
enables them to transform energy into motion, and therefore represent
systems far from thermodynamic equilibrium.  Active particles are
extensively studied from the experimental as well as from the
theoretical point of view.  Examples of corresponding studies range
from investigations of the dynamical behavior of individual units like
motile cells
\cite{schienbein.m:1993,teeffelen2008,selmeczi2008,bodeker2010,li_persistent_2008,Amselem,dilao_chemotaxis_2013,hauser15},
\notiz{over} macroscopic animals \cite{bazazi2010,RomRom15} and
artificial self-propelled particles
\cite{paxton2004,howse2007,ruckner2007,kumar2008,kudrolli2008,tierno2010,ke_motion_2010,takagi13},
to many body interactions and collective phenomena in many-particle
systems
\cite{vicsek1995,gregoire04,chate2006,buhl_disorder_2006,Bertin06,
  Ihle2011,Ihle2013,grossman2013,grossmann2016,hamid16,patch2017}.

We will concentrate on models describing the stochastic motion of an
individual active particle.  Such particles are often called
\textit{Active Brownian \notiz{P}articles} (ABP) because of their
similarities to normal Brownian motion
\cite{romanczuk.p:2012,schweitzer.f:1998,schweitzer.f:2007}.  Thus, at
large time scales both exhibit diffusive behavior, as calculations of
the mean square displacement confirm \cite{mikhailov}.  Moreover,
\notiz{a crossover} from an initial, ballistic to the final, diffusive
behavior have been reported \cite{romanczuk.p:2012,mikhailov,Peruani}
\notiz{as it is known for passive particles \cite{langevin1908} and
  for correlated random walks \cite{shigesada,okubo}}.  Differences
occur since an ABP is an object out of equilibrium \notiz{contrary} to
a normal Brownian particle whose motion corresponds to an equilibrium
situation \cite{ebeling.w:1999,erdmann_brownian_2000}.  The
fluctuating forces acting on ABP originate from the active nature of
the system, such as fluctuations of the propulsion, and not from
thermal noise\cite{romanczuk.p:2011,sevilla14}.

More precisely, we will report on a stochastic micro-swimmer moving
\notiz{with constant speed which is a special ABP
  \cite{romanczuk.p:2012}}.  Fluctuating forces act \notiz{only}
perpendicular to the instantaneous direction of motion. If these can
be described by white noise, the swimmer performs persistent ballistic
motion at small time scales and diffusive motion at larger scales,
similarly to a Brownian particle \cite{mikhailov}. For a Brownian
particle as well as for an ABP \notiz{including the micro-swimmer
  under consideration}, the ballistic regime is caused by the inertia
of the object. Therefore, the description of these two regimes, and of
the transition between them requires a formulation of the equations of
motion in phase space, including coordinates and velocity components,
or of the corresponding kinetic equations for joint probability
densities thereof.  \notiz{By} contrast, as well known, if only the
diffusive regime is of interest, the system can be modeled by a much
simpler overdamped dynamics \cite{Kramers,Becker52} for coordinates
alone, and the effects of inertia, describing the memory in the
velocity variable, can be eliminated.  \notiz{Then}, the velocity as a
variable follows instantaneously and without memory the forces acting
on the particle.

We will elaborate on such a procedure for the stochastic micro-swimmer
with constant speed. The variable introducing inertia in this model,
which has to be eliminated, is the orientation of the velocity. The
latter creates the persistent ballistic motion at small time scales
but does not have relevance for the diffusive regime \notiz{observed}
at larger time scales.  The consistent procedure of adiabatic
elimination of this orientational variable is the main topic of the
present work.

First, in \secref{brown}, we review the adiabatic elimination of the
velocity in the description of the normal Brownian particle
\cite{Becker52}.  It is presented for the Langevin equation as well as
\notiz{on} the level of a kinetic approach, which is similar to the
transport theory in gas dynamics \cite{Rumer80}. \notiz{A similar
  systematic reduction procedure in case of ABPs is nowadays still
  missing. Here, we fill this gap for the simple case of a
  micro-swimmer with constant speed as studied in
  \cite{haeggqwist,weber11,weber12,Thiel2012,noetl2017,mijalkov2013,volpe2014,geiseler2016,geiseler2016kramers,debnath,btenhagen,babel}
  and in many other applications}.  We will proceed in a similar way
as it is presented in \secref{brown} for a passive Brownian particle.
The corresponding new procedure is presented in \secref{active} again
on the level of the Langevin equation and also for the kinetic
approach. \notiz{In \secref{sec:concl} we summarize our findings}.

\section{Adiabatic Elimination for Brownian particles}
\label{brown}
The theory of Brownian motion is connected with such famous names as
Einstein \cite{einstein1905}, Langevin \cite{langevin1908} and
Smoluchowski \cite{smoluchowski1906}. It was a great success of the
developing statistical physics at the beginning of the 20$^\text{th}$
century being one of the first descriptions of dynamical fluctuations
using stochastic methods. Starting point of the description was the
formulation of a Chapman-Kolmogorov-like relation for the probability
density of the particle by Einstein. Soon later, Langevin claimed to
have found an infinitely simpler approach based on a stochastic
differential equation. Both, together with Smoluchowski, \notiz{can thus be}
identified as being pioneers in the developing tools for stochastic
dynamical problems.

\subsection{Adiabatic elimination I: Langevin equation}
Langevin's approach started from Newtonian dynamics in the presence of
friction and impacts described as noise force. Therefore, he
explicitly took into account inertia leading to the persistence of the
motion. \notiz{Einstein, by contrast, }obtained results for the
overdamped regime with high damping of the fluid or small masses of
the particles, where inertia can be neglected.

For the two-dimensional position vector $\mathbf{r}(t)\,=\,(x,y)$ and
the velocity $\mathbf{v}(t)\,=\,(v_x.\,v_y)$ of the particle, the
Langevin approach corresponds to equations of motion (in the nowadays
accepted notation) \cite{borkovec_1990}
\begin{equation}
  \label{eq:langevin1}
  \pt{\mathbf{r}}{t}\,=\,\mathbf{v},~~~~~~ m \, \pt{\mathbf{v}}{t}\,=\,-\gamma\,
  \mathbf{v}\,+\,\sqrt{2\gamma k_{\rm B}T}\, \boldsymbol{\xi} (t),
\end{equation}
where $m$ is the mass of the particles and \notiz{ $k_{\rm B}T$ is the
  thermal energy}. The two terms at the r.h.s. of the second equation
describe the interaction with the surrounding liquid. The first one
stands for the Stokes friction with coefficient $\gamma$ and the
second one models molecular agitation wherein
$\boldsymbol{\xi}=(\xi_x,\xi_y) $ is a Gaussian noise with vanishing
mean, having independent components $\xi_i$ and is $\delta$-correlated
in time, i.e.
\begin{equation}
\label{whitenoise}
\av{\boldsymbol{\xi}(t)}\,=\,0,~~~~~\av{\xi_i(t)\,\xi_j(t')}\,=\,\delta_{i,j}\delta\,(t-t')\,,~~~i,j\,=\,x,y.
\end{equation}
The prefactor of the noise term is called the noise intensity. It was
selected such that the particle's \notiz{velocity obey a Maxwellian
  distribution in equilibrium. The latter is established for $t \gg
  \tau_{\gamma}= m/\gamma$ where $\tau_{\gamma}$ is the relaxation
  time of the particle in the fluid. Hence, the mean squared velocity
  fulfills the equipartition requirement $m {v}_i^2/2 =k_B
  T/2,\,(i=x,y)$. The connection of the noise intensity with the
  friction coefficient is a static version of the
  fluctuation-dissipation relation.}

Integration of this system of equations yields the well known mean
squared displacement (MSD). In detail, for the initial condition
$\vec{r}(t=0)=\vec{r}_0=0$, it reads in \notiz{$d$ dimensions}
\cite{langevin1908}
\begin{equation}
\label{msd1}
\av{\mathbf{r}^2(t)}\,=\,\notiz{2\,d}\,\frac{k_{\rm
    B}T}{\gamma}\,\left[t\,+\,\frac{m}{\gamma}\left(\exp(-\frac{\gamma}{m}t)-1\right)\right]\,.
\end{equation}
For times $t \,\ll \,\tau_{\gamma}$ ballistic behavior is found.
Oppositely, for $t \,\gg \,\tau_{\gamma}$ diffusive behavior with a
linear growth of the MSD with respect to time is observed. The
corresponding time domains can be translated into distances traveled:
at distances smaller than $l_\gamma$ the behavior is ballistic, and
crosses over to diffusion at distances larger $l_\gamma$. The
crossover time is given by the relaxation time $\tau_{\gamma}$, and
$l_\gamma$ is known as the brake path,
$l_{\gamma}\,=\,\tau_{\gamma}\,\sqrt{k_{\rm B}T/m}\,$
\cite{Becker52,Hwalisz}.  In the long-time limit,
$t\,\gg\,\tau_{\gamma}$, when displacements are larger than the brake
path $l_{\gamma}$, this result coincide with Einstein's finding
\cite{einstein1905} and the diffusion coefficient reads
\notiz{approximately}
\begin{equation}
  \label{msd0}
  D_{\mbox{\tiny eff}}\,= \, \,\frac{1}{2\, d \,t}\, \av{\mathbf{r}^2(t)} \,\approx \,\frac{k_{\rm B}T}{\gamma}\,.
\end{equation}
\notiz{The approximation} is valid by assuming either the mass being
small or the friction being large. Therefore, this limit is called
nowadays ``overdamped''.

Being interested only in this coarse-grained long-time limit $t \gg
\tau_{\gamma}$ and $|{\mathbf r}|\gg l_{\gamma}$, one might reduce the
integration scenario by neglecting the inertia term in
\eqref{eq:langevin1}. To do this, one usually processes by collecting
the large items at the r.h.s. of the differential equation, i.e. one
formulates
\cite{borkovec_1990,Hwalisz,haken,gardiner,gardiner:82,ucna,lsg_talkner,sancho}
\begin{equation}
  \label{eq:langevin11}
  \pt{{\mathbf v}}{t}\, =\,-\frac {1}{\tau_{\gamma}} \left(\, \mathbf{v}\,-\,\sqrt{2\frac{ k_{\rm B}T}{\gamma}}\, \boldsymbol{\xi}(t)\right).
\end{equation}
Since $\tau_{\gamma}$ shall be small the items on the r.h.s. are large
supposed that the value of $k_{\rm B}T/\gamma $ remains finite.
\notiz{Then}, the l.h.s. of Eq.\eqref{eq:langevin11} becomes
negligible compared to each of the two items at the r.h.s.  which
therefore have to compensate each other. Thus effectively it holds
that
\begin{equation}
  \label{eq:elim1}
  \mathbf{v}\,=\,\sqrt{2\frac{ k_{\rm B}T}{\gamma}}\, \boldsymbol{\xi}(t).
\end{equation}
The validity is also proven since all higher moments of this
expression exist. Hence, as seen in Eq. \eqref{eq:elim1}, \notiz{for
  time scales larger} than $\tau_{\gamma}$ and with the assumed
coarse-graining, the velocity does not change smoothly but becomes
irregular white noise $\mathbf{v}(t) \propto \boldsymbol{\xi}(t)$. In
consequence, the autocorrelation function (ACF) of the velocity
becomes \notiz{$\delta$-like} reading
\begin{equation}
  \label{eq:autocorr0}
  K_{\mathbf{v},\mathbf{v}}(t\,-\,t')\,=\, \av{\mathbf{v}(t)\cdot\mathbf{v}(t')}\,=\, \notiz{2\, d}\,\frac{k_{\rm B}T}{\gamma}\,\delta(t-t')\,.
\end{equation}
Eventually, insertion of this velocity vector into the first equation
of \eq{eq:langevin1} defines the well-known overdamped dynamics for
the position of a Brownian particle \cite{borkovec_1990}
\begin{equation}
  \label{eq:langevin2}
  \pt{\mathbf{r}}{t} \,=\,\sqrt{2 \, \frac{k_{\rm B}T}{\gamma}}\, \boldsymbol{\xi}(t)\,,
\end{equation}
with $\boldsymbol{\xi}(t)$ being a Gaussian white noise with
properties defined in \eqref{whitenoise}.

Since the noise is additive we can simply integrate this equation and
obtain, after averaging over different realizations of the noise, for
the MSD 
\begin{equation}
\label{msd2}
\av{\mathbf{r}^2(t)}\,=\,\int_0^t\int_0^t  \,{\rm d}t\,{\rm d}t'\, K_{\mathbf{v},\mathbf{v}}(t-t').
\end{equation}
After taking the integral over the ACF $K_{\mathbf{v},\mathbf{v}}$
using \eqref{eq:autocorr0} one finds for the effective diffusion
coefficient exactly the r.h.s. of \eqref{msd0}.

\subsection{Adiabatic elimination II: Kinetic approach \label{sec:kinapp1}}

Elimination of velocities is also possible when
starting from the Fokker-Planck equation (FPE) for \notiz{the
  conditional probability density function
  $P(\mathbf{r},\mathbf{v},t|\mathbf{r}_0=0,\mathbf{v}_0,t_0)$, which
  is} the transition probability density to find the position and
velocity in an infinitesimal element $ {\rm d}\mathbf{r} {\rm
  d}\mathbf{v}$ around $\mathbf{r}$ and $\mathbf{v}$ of the phase
space at time $t$ having started at $t_0=0$ at the origin and with the
initial velocity $\mathbf{v}_0$, for simplicity. The corresponding FPE
reads
\begin{equation}
\label{fpe1}
\p{P}{t}\,=\,-\mathbf{v}\,\cdot\,\p{P}{\mathbf{r}}\,+\,\frac{\gamma}{m}\,\p{}{\mathbf{v}} \cdot \left(\,\mathbf{v}\, P\, + \,\frac{k_{\rm B}T}{m}\, \p{P}{\mathbf{v}} \right).
\end{equation}
\notiz{We would like to mention that already Kramers in his seminal work 
\cite{Kramers} found an elegant way to eliminate the velocity. 
Using the factorizing properties of the Fokker-Planck operator, 
he was able to derive the diffusion equation for the marginal probability density
$\rho(\mathbf{r},t)$ of the position of the Brownian particle by integrating 
the FPE \eqref{fpe1} over the velocity along an inclined straight line (see also 
the discussion in the excellent book by Becker \cite{Becker52}). Later, many other approaches have 
been formulated, including projection operator formalism \cite{haken,gardiner,gardiner:82}. 
To our best knowledge none of these approaches has been applied successfully to the problems 
of ABP so far. Therefore here we use a kinetic approach which is based on 
transport equations for the first three moments of the velocity. We first demonstrate this 
approach in application to a Brownian particle. Later on, the approach is successfully applied to the 
situation of the stochastic micro-swimmer with constant speed. }

The asymptotic solution for the marginal distribution of $\mathbf{v}$ is a 
Gaussian corresponding to the Maxwell distribution of velocities. 
In correspondence to the kinetic theory of gases, \notiz{the
  space- and time-dependent zeroth, first and second moments of the velocity are
  introduced. These} are the marginal probability density for the
position $\rho(\mathbf{r},t)$, the mean velocity with components
$u_i(\mathbf{r},t)$, and the variances $\braket{\delta v_i\,\delta v_j
}(\mathbf{r},t)$ of the deviations from the mean velocity $\delta
v_i\,=\,v_i\,-\,u_i$. In detail, they are defined
as\footnote{\notiz{Strictly speaking, these expressions are conditional moments. 
But, further on, we will omit for simplicity these conditions in the arguments of 
the pdf and of the moments, and include them via corresponding initial conditions of the moments.}}
\begin{eqnarray}
  \label{eq:moments1}
  &&\rho(\mathbf{r},t)\,= \,\int P(\mathbf{r},\mathbf{v},t){\rm d}\mathbf{v}
  \,,~~~~~~\rho(\mathbf{r},t)\,u_i(\mathbf{r},t)\, = \,\int v_i P(\mathbf{r},\mathbf{v},t){\rm d}\mathbf{v}\,,\\
  &&\rho(\mathbf{r},t)\,\braket{\delta v_i\,\delta v_j}(\mathbf{r},t)\, = \int (v_i-u_i(\mathbf{r},t))\,(v_j-u_j(\mathbf{r},t))\, P(\mathbf{r},\mathbf{v},t){\rm d}\mathbf{v}\,,~~~i,\,j\,=\,x,\,y.\nonumber
\end{eqnarray}
In kinetics these moments obey the transport equations which are
balance equations for the marginal density, momentum and temperature
\cite{Rumer80}. In our notation starting from the
FPE \eqref{fpe1} and by corresponding
multiplications and integrations we derive:
\begin{eqnarray}
\label{transport}
&&\p{\rho}{t}+\p{}{x_k}\rho\,u_k\,=0, ~~~~~~~~
\p{u_i}{t}+u_k\p{u_i}{x_k}= -\frac{\gamma}{m} u_i \notiz{-}\frac{1}{\rho}\p{\rho\braket{\delta v_i\,\delta v_k}}{x_k},\\\nonumber\\ 
&&\p{\braket{\delta v_i\, \delta v_j}}{t}\,+\,u_k\p{\braket{\delta v_i\, \delta v_j}}{x_k}\,= \\
&&~~~~~~~~~~~~~~=\, 2 \frac{\gamma}{m}\left(\frac{k_{\rm B}T}{m} \delta_{i,j}\,-\,
  \braket{\delta v_i\, \delta v_j}\right) \,-\,
\braket{\delta v_i \,\delta v_k}\p{u_j}{x_k}\, -\, \braket{\delta v_j \,\delta v_k} \p{u_i}{x_k} \nonumber
\end{eqnarray}
with $i,j,k=x,y$; the summation over the repeating indices is assumed. We close
the equations by neglecting third moments of deviations from the mean velocity, i.e. 
by assuming $\braket{\delta v_i \delta v_j \delta v_k}\approx 0$.

We now examine the asymptotic behavior of these quantities at times
longer than the relaxation time $t\,\gg\, \tau_{\gamma}$. In this
limit, the first bracket on the r.h.s. of the equation for the
variances becomes dominant since $\gamma/m=\tau_{\gamma}^{-1}$ is a
large parameter.  Therefore, we can neglect the substantial derivative
on the l.h.s. of the dynamics and the remaining items on the r.h.s.
yielding in zeroth order of small $\tau_{\gamma}$
\begin{equation}
  \label{eq:variance_as}
  \ \braket{\delta v_i\,\delta v_j}(\mathbf{r},t) \,\approx\,\frac{k_{\rm B}T}{m}\,\delta_{i,j}\,+\,\mathcal{O}(\tau_{\gamma})\,.
\end{equation}
The cross-correlations disappear, and the standard deviations for both
directions coincide, become homogeneous, and are time-independent.

Similarly, we proceed for the mean velocity. The substantial
derivative can be assumed to be small compared to the two terms on the
r.h.s.  of the equation for the mean velocity at large time scales.
The mean velocity will follow the evolution of the slowly developing
density.  In consequence, after insertion of \eqref{eq:variance_as}
the mean flux in the position space becomes
\begin{equation}
  \label{eq:velocity_as}
  \rho(\mathbf{r},t)\,u_i(\mathbf{r},t)\,=\,\frac{m}{\gamma}\,\p{ }{x_k} \,\rho\,\braket{\delta v_i\,\delta v_k}
  \approx\,- 
  \,\frac{k_{\rm B}T}{\gamma}\,\p{\rho}{x_i}. 
\end{equation}
The latter is a kind of Fick's law for the density flux
$\rho(\mathbf{r},t)\,u_i(\mathbf{r},t)$.  Thus, what remains as a
result of the adiabatic elimination of the mean velocity and of the
variances is the resulting continuity equation for the density which
reads
\begin{equation}
  \label{eq:density_as}
  \p{\rho}{t}\,=\,\,\frac{k_{\rm B}T}{\gamma}\,\left( \frac{\partial^2}{\partial x^2}\,+\,\frac{\partial^2}{\partial y^2}\right)\,\rho(\mathbf{r},t). 
\end{equation}
Eq.\eqref{eq:density_as} describes the normal diffusion of the
Brownian particles, and is the equation derived by Einstein in 1905
\cite{einstein1905}. It includes the effective diffusion coefficient
as defined in \eqref{msd0}.
Hence, by making effectively the same approximation as in the Langevin
approach, we derived the well-know equation for the probability
density $\rho(\mathbf{r},t)$ describing the diffusive behavior in the
overdamped regime.  Both approaches, the overdamped Langevin dynamics
\eqref{eq:langevin2} and the diffusion equation for the probability
density \eqref{eq:density_as} are equivalent models for the
description of the probabilistic motion of a Brownian particle. In the
next section we introduce the stochastic micro-swimmer with constant
speed. For the latter we will apply the same approach to formulate an
effective overdamped dynamics for the individual equation of motion as
well as for the kinetic equation for the probability density.

\section{Stochastic Micro-swimmer}
\label{active}
\subsection{Active Brownian particle with constant speed}
The new physics, making the particle to an active one, can be best
formulated on the level of the Newtonian law for a self-propelled
object. Here we consider the two-dimensional case \notiz{of an ABP};
the point particle is described by its position $\mathbf{r}(t)$ and
velocity $\mathbf{v}(t)$ at time $t$. Particles have a polarity which
points always along the current velocity and is represented by an unit
vector $\mathbf{e}_v(t)$ in the direction of velocity . In polar
coordinates (with the polar axis coinciding with the abscissa of the
Cartesian system) we have
\begin{equation}
  \label{eq:unit_vec}
  \mathbf{e}_v(t)\,=\,\left(\cos\phi(t),\sin\phi(t)\right), ~~~~~~~~~~~~\mathbf{e}_{\phi}(t)\,=\,\left(-\sin\phi(t),\cos\phi(t)\right)  
\end{equation}
whereby the orientation $\phi(t) \in [0,2\pi]$ is the angle
from the abscissa to the velocity vector. In consequence, the velocity vector reads
$\mathbf{v}(t)\,=\,v(t)\,\mathbf{e}_v(t)$, where $v(t)$ denotes the projection of the
velocity onto the polarity axes of the particle. Note, that the value
of $v(t)$ does not have a definite sign, it can also move backwards
with negative $v(t)$. The second vector $\mathbf{e}_{\phi}(t)$ defined in \eqref{eq:unit_vec} is the unit vector 
in the direction normal to the velocity, i.e. with angle shifted by $\notiz{+}\pi/2$. 

First, we consider the mean deterministic part of the propulsive force
which acts in the direction of the polarity axis, i.e. in the
direction $\mathbf{e}_v$ of the instantaneous velocity. For
simplicity, we assume a constant force $\gamma\,v_0$ as reported in
\cite{schienbein.m:1993}.  The second ingredient is given by the
stochastic forces. \notiz{In contrast} to normal Brownian motion in
equilibrium, we assume that these are generated by the propulsive
mechanism itself and act in both directions: parallel and
perpendicular to the polarity \notiz{of} the swimmer which coincides with the
instantaneous velocity.  Complementing this model with Stokes friction we
get the equations of motion in the form
\begin{equation}
  \label{eq:active1}
\pt{\mathbf{r}}{t}\,=\,\mathbf{v}\,=\,v\,\mathbf{e}_v,~~~~~  \pt{\mathbf{v}}{t}\,=\,\gamma\,v_0\,\mathbf{e}_v\,-\,\gamma\,\mathbf{v}\,+\,\sqrt{2\,D_v}\,\xi_v(t)\,\mathbf{e}_v\,+\,\sqrt{2\,D_{\phi}}\,\xi_{\phi}(t)\mathbf{e}_{\phi},
\end{equation}
\notiz{which is a simple model of an ABP}.  Here and later on the mass $m$ of the particle is set to unity and omitted. The dependence
on the mass can be easily reestablished \notiz{by} rescaling the noise intensity as
$D_{i}\, \to \,D_{i}/m^2\,,~~i\,=\,v,\,\phi$ and $\gamma \,\to
\gamma/m$. The noise sources $\xi_v$ and $\xi_{\phi}$ are independent 
with intensities $D_v$ and $D_{\phi}$; the corresponding forces act along the polarity axis
and perpendicular to it, respectively. We assume white Gaussian
noises with vanishing mean:
\begin{equation}
  \label{noise_a}
  \av{\xi_i(t)}\,=\,0,~~~~~\av{\xi_i(t)\,\xi_i(t')}\,=\,\delta_{i,j}\,\delta(t-t'),~~~~~i,j\,=\,v,\,\phi.
\end{equation}
It is important to point out that \notiz{in contrast} to normal
Brownian motion the stochastic forces are multiplicative being
dependent on the orientation.  We also mention that more complex
settings are discussed in the literature, see
\cite{RomRom15,romanczuk.p:2011}.

The acceleration on the l.h.s. can be decomposed along the two unit
vectors as
\begin{equation}
  \label{eq:velocity2}
  \pt{\mathbf{v}}{t}\,=\,\pt{v}{t}\,\mathbf{e}_v\,+\,v\,\pt{\phi}{t}\mathbf{e}_{\phi}.
\end{equation}
Comparison of the r.h.s. of \eqref{eq:active1} and \eqref{eq:velocity2}
leads to the equations of motion in the two perpendicular directions:
\begin{equation}
  \label{eq:velocity3}
  \pt{v}{t}\,=\, \gamma\,(v_0\,-\,v)\,+\,\sqrt{2\,D_v}\,\xi_v(t), ~~~~~\pt{\phi}{t}\,=\,\frac{1}{v}\,\sqrt{2\,D_{\phi}}\,\xi_{\phi}(t).
\end{equation}
Notabl\notiz{y}, the dynamics of the $v(t)$ is independent of the
orientation.

Further on, we will consider the model with constant speed, which
corresponds to the assumption $D_v=0$. Moreover, we assume $\gamma \to
\infty$ which allows to adiabatically eliminate the velocity component
$v(t)$ along the polarity axis. Making these two assumptions we obtain
\begin{equation}
\label{eq:speed}
\dot{\mathbf{r}}\,= \,\vec{v}\,=\,v_0\, \mathbf{e}_{v},~~~~~~\dot{\phi}\,=\,\frac{1}{v_0}\,\sqrt{2D_{\phi}}\,\xi_{\phi}(t).
\end{equation}
This is a frequently-used model for a micro-swimmer moving at a
constant
speed\cite{grossmann2016,mijalkov2013,volpe2014,geiseler2016,patch2017,debnath,weber11,weber12,noetl2017}.
\notiz{As demonstrated in this section, the stochastic micro-swimmer
  with constant speed is a special kind of an ABP
  \cite{romanczuk.p:2012}}. The dynamics is given by the variation of
the orientation \notiz{which is} due to the action of a stochastic
force. We emphasize that this dynamics is not overdamped since it
still has inertia, which is here expressed by the memory in the
orientation $\phi(t)$.

\subsection{Adiabatic elimination in  one dimension \label{sec:effD1D}}
Let us first consider the motion in projection on the $x$-axis of the
Cartesian system. The velocity of this projection is not a constant
since it changes with the orientation $\phi(t)$. The corresponding
system of stochastic differential equations reads
\begin{equation}
\label{eq:xmo1d}
\dot x\, =\,v_x\,= v_0\cos \phi,~~~~~~\dot \phi=\frac{1}{v_0}\sqrt{2D_{\phi}}\xi_{\phi}(t).
\end{equation}
Taking the derivative of the velocity in \eqref{eq:xmo1d} results in
the compact representation
\begin{equation}
  \dot v_x\,=\,-\,\sin \phi\,\sqrt{2D_{\phi}}\,\xi_{\phi}(t)\,.
\end{equation}
With multiplicative noise we have to declare the stochastic
integration rule which we will be used.  \notiz{In our physical model we interpret white noise as the} limit of a short time correlated noise, i.e. the
noise correlation time is small compared to all \notiz{relevant} time scales of
the model. Then it is straightforward to formulate the problem by
means of the Stratonovich calculus
\cite{sancho,strato,vankampen,sokolov}\notiz{.} Therein, variable
transformations can be performed also without additional Ito-terms.

\notiz{This} yields the stochastic differential for the velocity increment
\begin{equation}
\label{eq:dvx}
  {\rm d} {v_x}=-v_0\sin\left(\phi+\frac{{\rm d}\phi}{2}\right)\,{\rm d} \phi 
\end{equation}
where the angular increment is a scaled Wiener-process:
\begin{equation}
\label{eq:dphi}
  {\rm d}\phi=\frac{1}{v_0}\sqrt{2D_{\phi}}\,{\rm d} W_t\,
\end{equation}
with first and second moments given by $\braket{{\rm d} W_t}\,=0$,
$\braket{{\rm d} W_t\,{\rm d} W_{t'}} \,=\,0$, for $t \ne t'$, and
$\braket{({\rm d} W_t)^2}\,=\,{\rm d}t$.

Using trigonometric theorems, equation \eqref{eq:dvx} can be rewritten as
\begin{equation}
  {\rm d} v_x\,=\,-\,v_0 \left[\sin \phi \,\cos\left(\frac{{\rm d}\phi}{2}\right)\,+\,\cos \phi\,\sin\left(\frac{{\rm d}\phi}{2}\right)\right]{\rm d} \phi\,.
\end{equation}
Since the angular increment is infinitesimal, ${\rm d}
\phi\ll 1$, insertion of \notiz{equation \eqref{eq:dphi}} gives
\begin{equation}
  {\rm d} v_x\, =\,-\,\sqrt{2D_{\phi}}\,\sin \phi \,{\rm d} W_t\,-\,\frac{D_{\phi}}{v_0}\,\cos \phi\,({\rm d} W_t)^2\,.
\end{equation}
\notiz{Note} that the orientation $\phi(t)$ in this expression is
statistically independent of the increment of the Wiener process ${\rm
  d} W_t$ being generated after $t$.  Then we use the fact that $\cos
\phi = v_x/v_0$ and the property of the Wiener process that the
\notiz{non}-averaged squared increment behaves as ${(\rm
  d}W_t)^2\,=\,{\rm d}t\,+\, \mathcal{O}({\rm d}t^{3/2})$
\cite{strato2}.  \notiz{This} yields for the stochastic differential
in first order of ${\rm d} W_t$ and ${\rm d} t$
\begin{equation}
  {\rm d} v_x\, =\,-\,\sqrt{2D_{\phi}}\,\sin \phi\,{\rm d} W_t\,-\,\frac{v_x}{v_0^2}\,D_{\phi}\,{\rm d} t\,.
\end{equation}
From this differential form we can return to the stochastic
differential equation which results in the Ito equation
within the Stratonovich calculus reading
\begin{eqnarray}
\label{eq:overdamped}
  \dot v_x\,=\,-\frac{D_{\phi}}{v_0^2}v_x\,-\,\sin \phi\,\sqrt{2D_{\phi}}\,\xi_{\phi}(t)\,. 
\end{eqnarray}
This equation will allow for an adiabatic elimination of the velocity
components $v_x$. The characteristic time scale
$\tau_{\phi}\,=\,v_0^2/D_{\phi} $ describes the crossover between the
ballistic and the diffusive behavior. On longer time scales $t\,\gg
\tau_{\phi}$ the parameter $\tau_{\phi}$ acts as a small parameter.
Rewriting \eqref{eq:overdamped} as
\begin{eqnarray}
\label{eq:overdamped11}
\dot v_x\,=\,-\frac{1}{\tau_{\phi}}\,\left( v_x\,+\,\sin \phi\,\sqrt{2 \frac{v_0^4}{D_{\phi}}}\,\xi_{\phi}(t)\right)\,\notiz{,} 
\end{eqnarray}
the r.h.s is much larger and one can expect a fast relaxation of the
velocity component by setting effectively $\dot{v}_x=0$.  Just like
for the Brownian motion, see Eq.\eqref{eq:langevin2}, the two items on
the r.h.s. compensate. Therefore, the velocity becomes white noise
\begin{equation}
  \label{eq:overdamped5}
  \dot x\,=\,v_x\,=\,-\,\sin \phi\,\sqrt{2 \frac{{v_0^4}}{D_{\phi}}}\,\xi_{\phi}(t)\,. 
\end{equation}
This equation formulates the overdamped dynamics of the $x$-
components of a micro-swimmer with constant speed. It \notiz{links the velocity component 
to an instantaneous} orientation $\phi(t)$. It has to be
noticed that this random orientation is independent from the
instantaneous value of the noise $\,\xi_{\phi}(t)$ at the present time
$t$ arising from the increment of the Wiener process ${\rm d}W_t$
generated after $t$.

\notiz{By setting $\dot{v}_x\,=\,0$ at times $t \gg \tau_{\phi}$ the
  angular velocity $\dot{\phi}$ consequently vanishes as well. This is
  seen in Eq.  \eqref{eq:dvx}. Hence the history of the orientation
  $\phi(t)$ as given by Eq. \eqref{eq:dphi} is lost. The value of
  $\phi$ in Eq.  \eqref{eq:overdamped5} is given} by a series of
independent random numbers generated from the probability density of
$P(\phi,t)$. \notiz{ The latter, as shown in the Appendix
  \ref{sec:anglePDF}, becomes} stationary and homogeneously distributed in
$[\,0\,, 2 \pi\,]$ at the considered time scales
$t\,\gg\,\tau_{\phi}$, i.e. $P(\phi,t) \to P^0(\phi)= 1/2\pi$.


From Eq.\eqref{eq:overdamped5} one also sees that there is no motion
for $\phi\,=\, 0\,; \pi$. This results from the fact that the angular
noise acts perpendicular to the current orientation. Correspondingly,
the increments of \notiz{horizontal} position become largest for
$\phi\,=\, \pi/2\,; 3\pi/2$. This is confirmed by the velocity ACF
given by
\begin{eqnarray}
\label{eq:auto1}
\braket{v_x(t)v_x(t')}\,=\,\frac{2 v_0^4}{D_{\phi}}\sin^2(\phi)\,\delta(t-t')\,
\end{eqnarray}
as a function of $\phi$. It vanishes along the $x$-axis and becomes
maximal with perpendicular orientation.

\subsection{Adiabatic elimination in two dimensions
  \label{sec:effD2D}} Now we elaborate on the effective overdamped
dynamics in two dimensions. We proceed in a similar way as in the
previous section. However, the situation is a bit more complicated
since the motion in projections on both axes is correlated due to the
action of a single noise source.

Taking the derivatives of the vector
${\mathbf{v}}\,= \,v_0 \mathbf{e}_{v}$ one gets the equations of motion which read
\begin{equation}
  \label{eq:2d1}
\dot{\mathbf{v}}\,=\,v_0 \mathbf{e}_{\phi}\,\dot{\phi}\,=\,\mathbf{e}_{\phi}\,\sqrt{2D_{\phi}}\,\xi_{\phi}(t),
\end{equation}
where we have inserted the corresponding dynamics for the orientation
$\phi$ as in Eq.\eqref{eq:xmo1d}. Thus, as assumed in the model, the
acceleration acts perpendicular to the velocity. We note that both
vector components have the same acting noise at time $t$ \cite{weber11,weber12,noetl2017}.
Applying the Stratonovich calculus gives for the increments of both
velocity components $v_x(t)\,=\,v_0\,\cos(\phi(t))$ and
$v_y(t)\,=\,v_0\,\notiz{\sin}(\phi(t))$ in lowest order in ${\rm d}W_t$
and ${\rm d} t$
\begin{equation}
  \label{eq:2d2}
  {\rm d} v_x\,=\,-\frac{v_x}{v^2_0}\,D_{\phi}\,{\rm d} t\,-\,\sin{\phi}\,\sqrt{\notiz{2\,}D_{\phi}}\,{\rm d} W_t\,,~~~~~~~{\rm d} v_y\,=\,-\frac{v_y}{v^2_0}\,D_{\phi}\,{\rm d} t\,+\,\cos{\phi}\,\sqrt{\notiz{2\,}D_{\phi}}\,{\rm d} W_t\,. 
\end{equation}
Returning to the stochastic differential equations and assigning again
$\tau_{\phi}=v_0^2/D_{\phi}$, we get:
\begin{equation}
  \label{eq:2d3}
  \dot{\mathbf{v}}\,=\,-\,\frac{1}{\tau_\phi}\,\left( \mathbf{v} \,-\, \sqrt{2 \frac{v_0^4}{D_{\phi}}}\,\mathbf{e}_{\phi}\,\xi_{\phi}(t)\right)
\end{equation}
First, we underline that in \eqref{eq:2d3} the noise $\xi_{\phi}(t)$
and the orientation $\phi$ are independent from each other since the
increment of the Wiener process in the interval $[t,t+{\rm d}t]$ is
independent from its value at time $t$. Secondly, it is worth \notiz{pointing}
out that the noise $\xi_{\phi}(t)$ and the current orientation
$\phi(t)$ have identical values \notiz{for} both vector
components.

Equation \eqref{eq:2d3} allows again for the adiabatic elimination of
the velocity as a variable for $t \,\gg \, \tau_{\phi}$ and for
spatial increments larger \notiz{than} $l_{\phi}$. Under those assumptions
$\tau_{\phi}$ is again a small parameter forcing a fast relaxation of
the velocity vector.  We put the l.h.s. of the equation \eqref{eq:2d3}
to zero and solve the r.h.s. for the velocity $\mathbf{v}(t)$. Just
like in our previous discussion this assumption has as a consequence
that, according to Eq.\eqref{eq:2d1}, any history of $\phi(t)$
disappears, and the values of $\phi(t)$ become independent. As shown
in the Appendix \ref{sec:anglePDF}, the wrapped $\phi$ is homogeneously
distributed in the interval between $0$ and $2\pi$.

Hence, for times $t \,\gg \, \tau_{\phi}$ the velocity transforms into
scaled white noise
\begin{equation}
  \label{eq:2d4}
  \dot{\mathbf{r}}\, =\,\mathbf{v}\, =\, \sqrt{\frac{2\,v_0^4}{D_{\phi}}}\,\mathbf{e}_{\phi}\,\xi_{\phi}(t)\,.
\end{equation}
This vector equation is the sought-after overdamped dynamics of the
two-dimensional micro-swimmer \notiz{\footnote{We acknowledge the unknown referee 
for giving the hint that the found overdamped dynamics of a micro-swimmer with constant 
speed is independent of the dimensionality of the motion.}}.  Both components of the velocity are
correlated due to the same random orientation $\mathbf{e}_\phi(t)$
given by the value of $\phi$ at time $t$ and due to the same noise
$\xi_{\phi}(t)$ defining the strength of the torque acting on the
particle.  The $x$-component vanishes if the random orientation
$\phi(t)$ points along the $x$-direction.  Respectively, the
$y$-component vanishes if the orientation is parallel to the
$y$-direction. The reason is that in the model, no change in the speed
is allowed, and the noise is perpendicular to the polarity axis
showing in the velocity direction.

Scalar multiplication of the r.h.s. of \eqref{eq:2d3} with itself and
averaging over different realizations of noise gives the memoryless
velocity ACF:
\begin{equation}
\label{eq:corr1}
\braket{\mathbf{v}(t)\,\cdot \mathbf{v}(t')}\,=\,\frac{2 v_0^4}{D_{\phi}}\,\delta(t-t')\,.
\end{equation}
This expression is independent \notiz{of} the \notiz{current} value of the orientation
$\phi(t)$ which expresses the homogeneous distribution of the
orientation. Nevertheless\notiz{,} the contributions of the different
components still depend on the random orientation.
\begin{equation}
  \label{eq:corr11}
  \braket{v_x(t)\,v_x(t')}\,=\,\frac{2 v_0^4}{D_{\phi}}\,\sin^2(\phi)\,
  \delta(t-t')\,,
  ~~~~~~  \braket{v_y(t)\,v_y(t')}\,=\,\frac{2 v_0^4}{D_{\phi}}\,\cos^2(\phi)\,\delta(t-t')\,.
\end{equation}
For the cross-correlation function of the two velocity components we
get
\begin{equation}
\label{eq:corr2}
\braket{v_x(t)\,v_y(t')}\,=\,-\,\frac{2 v_0^4}{D_{\phi}}\,\sin(\phi(t))\,\cos(\phi(t'))\,\delta(t-t').
\end{equation}
Properties of \eqref{eq:corr11} and \eqref{eq:corr2} define the mean
squared increments per unit time. From this one can derive the
corresponding two-dimensional Smoluchowski equation for the
probability density $\rho(x,y,t)$ corresponding to \eqref{eq:2d4}.
It reads
\begin{equation}
  \label{eq:smolu1}
  \p{\rho}{t}\,=\,\frac{v_0^4}{D_{\phi}}\,\left( \,-\,\p{}{x}\,\sin \phi+\,\p{}{y}\,\cos \phi\right)^2\,\rho.
\end{equation}

Alternatively, the mean square displacement can be easily calculated
by using the connection between the MSD and the ACF \eqref{msd2}.
Insertion of \eqref{eq:corr1} and taking the initial condition again
in the origin $\mathbf{r}_0=0$ results in
\begin{equation}
  \label{eq:msd1}
  \braket{\mathbf{r}^2(t)}\,=\,2\,\frac{v_0^4}{D_{\phi}}\,t\,.
\end{equation}
The effective diffusion coefficient defined in \eqref{msd0} becomes
for the two dimensional case of \notiz{an} overdamped stochastic micro-swimmer
with constant speed
\begin{equation}
  \label{eq:eff_diff}
  D_{\mbox{\tiny{eff}}}\,=\,\frac{v_0^4}{2\,D_{\phi}}\,.
\end{equation}
It is the awaited result in the long time limit as first obtained by
Mikhailov and Meink\"ohn \cite{mikhailov}.

\subsection{Average over random orientations \label{sec:aver}}
We derived a memoryless description for the stochastic micro-swimmer
with constant speed and obtained normal diffusive behavior with the
awaited diffusion coefficient. Some remarkable differences from the case of an
overdamped Brownian particle must however be stressed.  For example, displacements along the $x$ and
$y$-axes are still correlated, as expressed by the cross-correlation
function \eqref{eq:corr2}. 

\begin{figure}[ht]
\centerline{
  \includegraphics[width=0.32\textwidth]{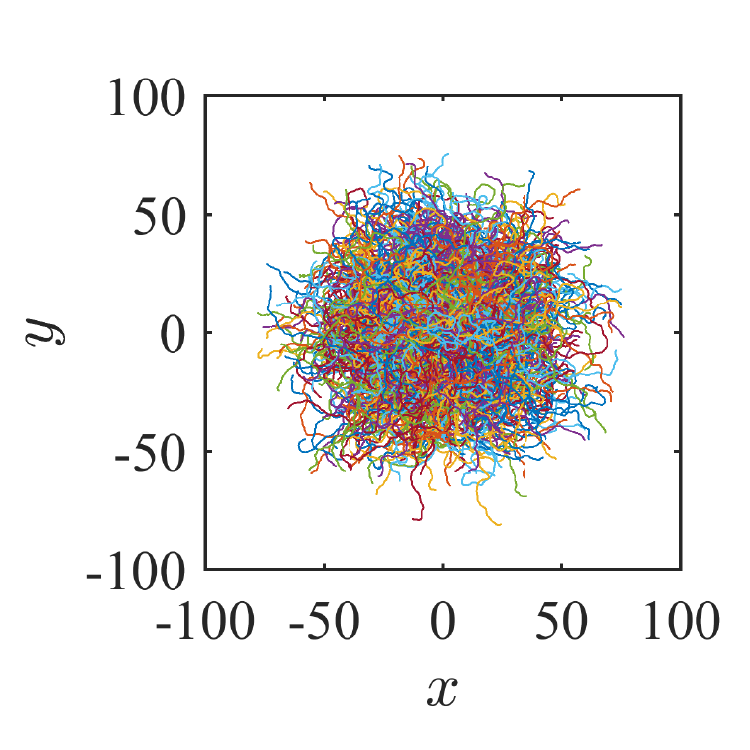}~\includegraphics[width=0.32\textwidth]{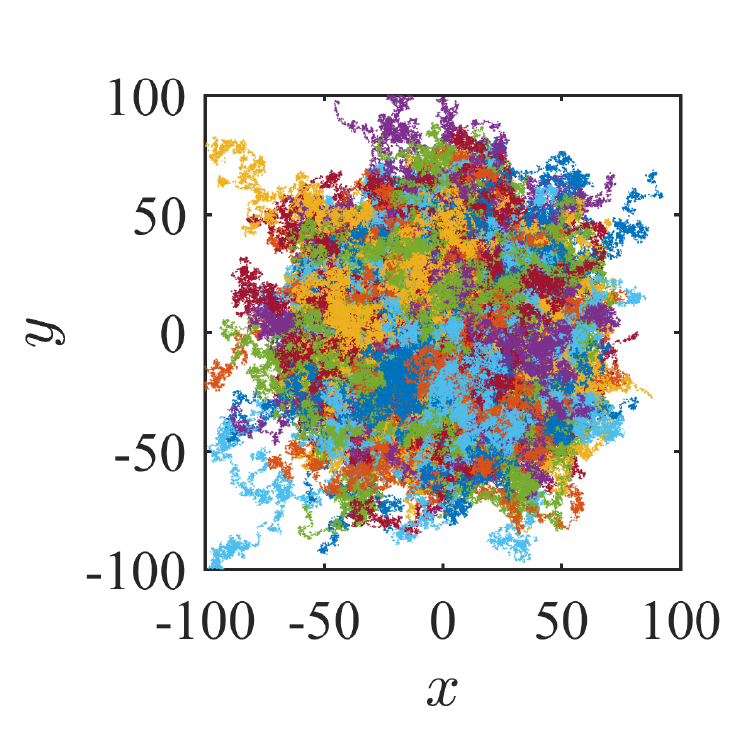}~\includegraphics[width=0.32\textwidth]{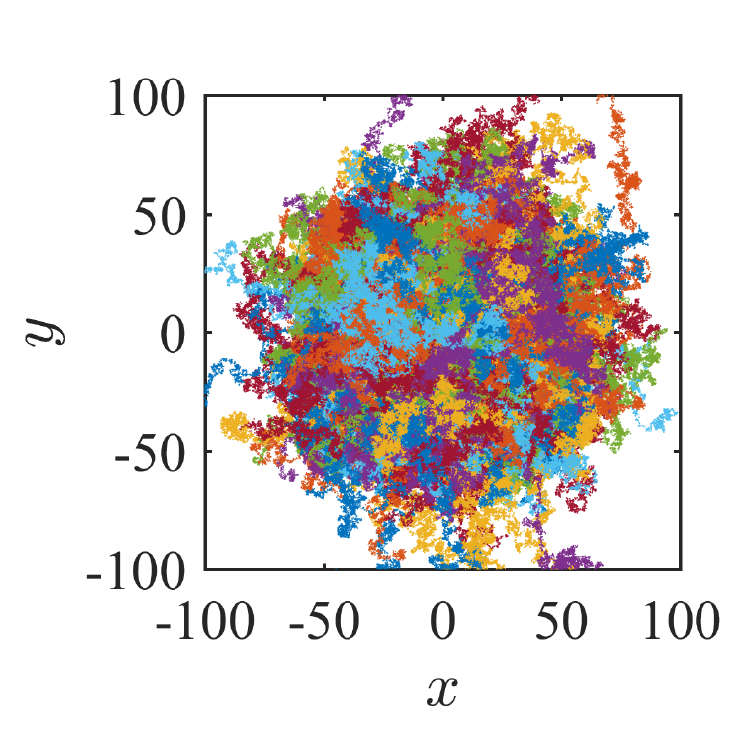}}
\caption{Trajectories of $N= 1000$ simulated particles in two
  dimensions ($t=10\,\tau_{\phi}$). left: Micro-swimmer with inertia,
  Eq.\eqref{eq:speed}; middle: Swimmer with random orientation and one
  stochastic force, Eq.\eqref{eq:2d4}; right: Swimmer with two
  independent stochastic forces, Eq.\eqref{eq:2d6}.  \notiz{Different colors show different trajectories.} Parameters: $N=
  1000; v_0=1; D_{\phi}=0.1; dt=0.02$. }
 \label{traj_abp}
\end{figure}

To describe the \notiz{uncorrelated} diffusive behavior one can make another reductive
step using the fact that the orientations are homogeneously distributed at large time
scales $t\gg \tau_{\phi}$ as \notiz{mentioned} several times \notiz{before}. \notiz{When averaging} over the \notiz{uniformly} distributed orientations, both
velocity components become \notiz{uncorrelated}. Averaging \notiz{Eq. \eqref{eq:corr2}}
results in
\begin{eqnarray}
  \label{eq:2d5}
  \av{\av{v_x(t)v_y(t')}}_{\phi}\,=\,0\,.
\end{eqnarray}
\notiz{If we assume the distributions of $v_x$ and $v_y$ at $t\gg
  \tau_{\phi}$ to be Gaussian, the components of velocity get to be
  statistically independent} and identically distributed:
\begin{equation}
  \label{eq:2d55}
  \av{\av{v_x(t)v_x(t')}}_{\phi} \,=\,\av{\av{v_y(t)v_y(t')}}_{\phi}\,=\,\frac{v_0^4}{D_{\phi}}\,\delta(t-t')\,.  
\end{equation}
Therefore, as a consequence of the elimination procedure and of the
averaging over the orientation, both components of the velocity can be
modeled as being forced by two independent white noise sources
\begin{equation}
  \label{eq:2d6}
  \dot x\,=\,\frac{v_0^2}{\sqrt{D_{\phi}}}\,\xi_{x}(t)\,,~~~~~~~~\dot y\,=\,\frac{v_0^2}{\sqrt{D_{\phi}}}\,\xi_{y}(t)
\end{equation}
as defined in \eqref{whitenoise}. Resulting trajectories, \notiz{as
  presented in Fig.\ref{traj_abp}c are statistically indistinguishable
  from the description in Sec.\ref{sec:effD2D} which sample paths are
  shown in Fig.\ref{traj_abp}b}.  \notiz{The independence of the
  components of the velocity also follows immediately from the
  following discussion: } The stochastic process \eqref{eq:2d6} can be
described by the probability density $\rho(x,y,t|x_0,y_0,t_0)$ which
obeys the two-dimensional diffusion equation
\begin{equation}\label{eq:smolu2}
  \frac{\partial}{\partial t} \rho\,=\,\frac{v_0^4}{2 D_{\phi}}\,\left(\frac{\partial^2}{\partial x^2}\,+\,\frac{\partial^2\,}{\partial y^2}\right)\,\rho\,,
\end{equation}
which can be obtained by averaging \eqref{eq:smolu1} over the random
orientation. From \eqref{eq:smolu2}, we can read again the effective
diffusion coefficient as presented in \eqref{eq:eff_diff}.  \notiz{The
  independence of velocity components follows from variables'
  separation in this equation under which the PDF factorizes.}  We
underline, that the crossover time to the diffusive motion with an
effectively overdamped dynamics coincides with relaxation time of the
velocity components which is $\tau_{\phi}\,=\,v_0^2/D_{\phi}$. It is
just the time where the initially inhomogeneous orientations are
forgotten, and the orientation becomes homogeneously distributed.
Beyond this time the motion of the micro-swimmer with constant speed
becomes statistically indistinguishable from the motion of a Brownian
particle.  The difference is that the diffusion coefficient scales
counter-intuitively with the intensity of the applied angular noise
behaving as $ D_{\mbox{\tiny{eff}}}\,\propto 1/D_{\phi}$ and
decreasing when this intensity increases, whereas the normal diffusion
enhances with increasing noise.

\begin{figure}[ht]
\begin{minipage}{0.59\textwidth}
   \centerline{\includegraphics[width=1.\textwidth]{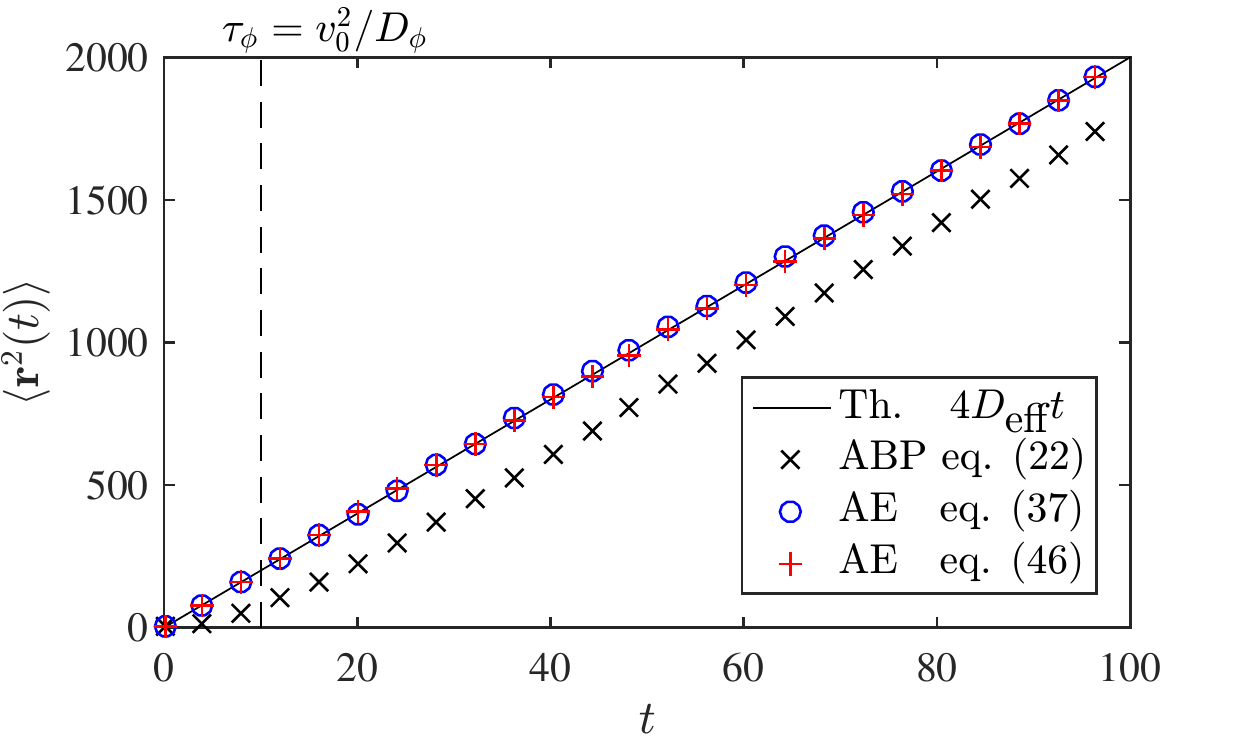}}
\end{minipage}
\begin{minipage}{0.29\textwidth}
  \caption{Simulated MSD in two dimensions: Comparison of the three
    different dynamics as labeled. The dashed line indicates the
    crossover time between the ballistic and the diffusive motion.
    Parameters are: $N= 10000; v_0=1; D_{\phi}=0.1; dt=0.02$.}
 \label{msd_abp}
\end{minipage}
\end{figure}

In Fig.\ref{traj_abp} we show sample paths for the two dimensional
stochastic micro-swimmer with constant speed. The left graph shows the
micro-swimmer with inertia \eqref{eq:speed}, the middle graph the
memoryless dynamics with random orientations according to
\eqref{eq:2d4}. In the right graph trajectories are presented with two
independent noise sources, corresponding to \eqref{eq:2d6}. Both
approximations exhibit the same diffusive behavior where the diffusion
coefficients converges to the asymptotic value of the case with
inertia as presented in Fig.\ref{msd_abp}. The mean square
displacement of the latter one is smaller compared to the overdamped
dynamics due to the initial ballistic growth.

\subsection{Adiabatic elimination: Kinetic approach\label{sec:kinapp2}
}
In this section we derive the Smoluchowski equation of the
micro-swimmer with constant speed following the kinetic approach in
\secref{sec:kinapp1}. The starting FPE for the joint \notiz{transition probability
density $P(\mathbf{r},\phi,t|\mathbf{r}_0=0,\phi_0,t_0)$ of the position $\mathbf{r}$ 
and orientation $\phi$ at time $t$ of the micro-swimmer with constant
speed which has been started at $t_0$ in $\mathbf{r}_0$  with orientation $\phi_0$. It }reads 
\notiz{\footnote{Again we omit the condition in the transition pdf and formulate corresponding initial conditions for the moments.}}:
\begin{equation}
  \label{eq:fpe5}
  \p{P}{t}\,=\,-\,v_0\,\p{}{\mathbf{r}}\, \cdot \,\mathbf{e}_{v}\,P\,+\,\frac{D_{\phi}}{v_0^2}\,\frac{\partial^2}{\partial {\phi}^2}\,P .
\end{equation}
We derive again transport equations for the zeroth, first and
second moments of the orientation which are
\begin{eqnarray}
  \label{eq:moment5}
  &&\rho(\mathbf{r},t)\,= \,\int P(\mathbf{r},\phi,t)\,{\rm d}\phi\,,~~~~~~\rho(\mathbf{r},t)\,u_i(\mathbf{r},t)\, = \,v_0\,\int\,\mathbf{e}_{v_i}\,  P(\mathbf{r},\phi,t)\,{\rm d}\phi\,,\\
  &&\rho(\mathbf{r},t)\,\braket{\delta v_i\,\delta v_j}(\mathbf{r},t)\, = \,\int (v_0\,\mathbf{e}_{v_i}\,-\,u_i(x,t)) (v_0\,\mathbf{e}_{v_j}\,-\,u_j(x,t))\,P(\mathbf{r},\phi,t)\,{\rm d}\phi.\nonumber 
\end{eqnarray} 
Multiplication with zeroth, first and second powers of $\cos(\phi)$
and $\sin(\phi)$ and integration over $\phi$ results in the three
transport equations
\begin{eqnarray}
\label{eq:moment6}
&&\p{\rho}{t}+\p{}{x_k}\rho\,u_k\,=0, ~~~~~~~~
\p{u_i}{t}+u_k\p{u_i}{x_k}= -\frac{D_{\phi}}{v_0^2} u_i -\frac{1}{\rho}\p{\rho\braket{\delta v_i\,\delta v_k}}{x_k},\\\nonumber\\ 
&&\p{\braket{\delta v_i\, \delta v_j}}{t}\,+\,u_k\p{\braket{\delta v_i\, \delta v_j}}{x_k}\,= \\
&&~~~~~~~~~~~~~~=\, 2 \frac{D_{\phi}}{v_0^2}\left(v_0^2 \delta_{i,j}\,-\,u_i\,u_j\,-\,
  2\,\braket{\delta v_i\, \delta v_j}\right) \,-\,
\braket{\delta v_i \,\delta v_k}\p{u_j}{x_k}\, -\, \braket{\delta v_j \,\delta v_k} \p{u_i}{x_k}.\nonumber
\end{eqnarray}
In these equations we can eliminate the first and second \notiz{moments} at times larger
\notiz{than} the relaxation time $t \,\gg \,\tau_{\phi}$ assuming large noise
$D_{\phi}$ or small velocities $v_0$. Since in this case the r.h.s of the
corresponding balance equations are large, the substantial derivatives in their l.h.s.
can be set to zero. From the equation for the variance $\braket{\delta
  v_i \, \delta v_j}$ we get
\begin{equation}
  \label{eq:moment7}
  \braket{\delta v_i \, \delta v_j}\, \approx \, \frac{1}{2}\left(v_0^2\delta_{i,j} \,-\,u_i\, u_j\right)\,+\,O(\tau_{\phi})\,, ~~~~~~~i,j\,=\,x,y.
\end{equation}
where we assumed small variations of the mean velocities \notiz{$\partial
u_i/\partial x_k$ which would contribute to the variances in $O(\tau_{\phi})$ }.  The fact that the variance depends on the
components of the mean velocity expresses the conservation of the mean
kinetic energy (the kinetic energy of a particle moving at a constant speed is constant). This \notiz{is clearly expressed by}
\begin{equation}
  \label{eq:kinetic}
  u_x^2\,+\,u_y^2\,+\,\braket{(\delta v_x)^2}+\,\braket{\delta v_y)^2}\,=\,v_0^2. 
\end{equation}
The approximation \eqref{eq:moment7} obeys this conservation law on the average. It
represents a kind of equipartition of the kinetic energy between both
degrees of freedom: each degree on the average obtains $v_0^2/2$. For the
mixed components one gets correspondingly on the average $\,u_x\,
u_y\,+\, \braket{\delta v_x \, \delta v_y}\, \approx \,0$.

Setting the l.h.s. to zero in the balance for the mean velocity
results in
\begin{equation}
  \label{eq:moment8}
  \rho\,u_i\,\approx\, -\,\frac{v_0^2}{D_{}\phi}\, \p{}{x_k}\,\braket{\delta v_i\,\delta v_k} \,\rho.
\end{equation}
Insertion of Eq.\eqref{eq:moment7} into Eq.\eqref{eq:moment8} and of the resulting expression into the continuity equation \notiz{finally gives}
\begin{equation}
  \label{eq:moment9}
  \p{\rho}{t}=\frac{\tau_{\phi}}{2}\, v_0^2\,\left(\frac{\partial^2}{\partial x^2}+\frac{\partial^2}{\partial y^2}\right)\,\rho-\frac{\tau_{\phi}}{2}\left(\p{}{x}u_x+\p{}{y}u_y\right)^2\rho. 
\end{equation}
Eventually, we obtained the equation for the marginal probability
density of the position $\rho(\mathbf{r},t)$ which is a valid
approximation to characterize the dynamics of the micro-swimmer. So
far still containing the mean velocity, it is not a closed
description. Both terms are of linear order in $\tau_{\phi}$. However\notiz{,}
we have to point out that the first one is multiplied by $v_0^2$
whereas the second one contains the components of the mean velocity
$u_x,u_y$. \notiz{Though the mean velocities scale with the speed $v_0$, their values are much smaller than the constant speed if a sufficiently strong noise acts on the orientation}.  Neglecting \notiz{in this situation consequently the
second term in Eq. \eqref{eq:moment9} yields} the two\notiz{-}dimensional diffusion equation
\eqref{eq:smolu2} with the constant diffusion coefficient
\eqref{eq:eff_diff}.  Its solution gives the \notiz{expected} approximation for
$\rho(\mathbf{r},t)$. \notiz{Otherwise, as always in cases where the motion is bounded by a maximal velocity, the diffusion approximation breaks down at the wings of the distribution. It happens in our case, when the mean velocity gets of the order of $v_0$.}


\section{Conclusions \label{sec:concl}}
We have been concerned with a stochastic micro-swimmer with constant
speed \notiz{and} random orientation. \notiz{This special type of an ABP}
performs ballistic motion at time scales smaller than the crossover
time $\tau_{\phi}$ and exhibits normal diffusion at larger time
scales. Our main purpose was to find an approach to an adiabatic
elimination of the orientation of the velocity as a persistent
variable, and to formulate a memoryless dynamics for the position of
the particle in the diffusive regime.  In analogy to the known theory
of a normal Brownian particle \notiz{which was revisited in
  Sec.\ref{brown}, we have derived in Sec.\ref{active} } the
overdamped dynamics of the projection of the motion onto a given
direction \eqref{eq:overdamped5} and also for the full two-dimensional
case, Eq.\eqref{eq:2d4}. On the basis of stochastic differential
equations \notiz{we elaborated systematic} elimination procedure. In addtion we propose a kinetic approach where we eliminate systematically the first and second spatio-temporal moments of the velocity and remain with the marginal density $\rho(\mathbf{r},t)$ to find the particle at time $t$ at position $\mathbf{r}$.
\notiz{In both approaches resulting equations} still reflect important microscopic
properties of the micro-swimmer as, for instance, the conservation of
kinetic energy. Only after averaging over random orientations the
particles loses this property and the description becomes similar
\notiz{to that of a} normal \notiz{overdamped} Brownian particle.

\vspace{1cm}
\noindent
\textbf{Acknowledgements} \\The authors cordially congratulate Ulrike
Feudel on occasion of her birthday and thank for a long-lasting
friendship. This work was supported by the Deutsche
Forschungsgemeinschaft via IRTG 1740.


\appendix

\section{Angular Probability Density Function \label{sec:anglePDF} }
The marginal probability density $P(\phi,t|\phi_0,t_0)$ of the orientation of
active Brownian particles obeys the Smoluchowski equation. The latter
can be obtained by integrating the FPE \eqref{eq:fpe5} over
the position vector $\mathbf{r}$ which results in
\begin{eqnarray}
  \frac{\partial P(\phi,t|\phi_0,t_0)}{\partial t}=\frac{D_{\phi}}{v_0^2}\frac{\partial^2 P(\phi,t|\phi_0,t_0)}{\partial \phi^2}.
\end{eqnarray}
For this parabolic equation the time-dependent solution for
$2\pi$-periodic boundaries is known \cite{Tikhonov}
\begin{equation}
  \label{orient1}
  P(\phi,t|\phi_0,0)\,=\,\frac{1}{\pi}\left(\frac{1}{2}+\sum_{n=1}^{\infty}\cos[n(\phi-\phi_0)]e^{-n^2\frac{D_{\phi}}{v^2_0} t}\right).
\end{equation}
For $t\,\gg \,\tau_{\phi}\notiz{=v_0^2/D_{\phi}}$ 
the probability spreads homogeneously over the $[0,2\pi ]$ interval:
\begin{equation}
  P^0(\phi)\,=\,\frac{1}{2\pi}.
\end{equation}
The characteristic time scale of the system is the relaxation \notiz{time $\tau_{\phi}$ after which
the first Fourier mode decays}.

\section{Mean squared displacement of the stochastic micro-swimmer
  with constant speed\label{sec:kubo} } 
  In 1997 Mikhailov and Meink\"ohn \cite{mikhailov}
derived the effective diffusion constant for the micro-swimmer with
constant speed using the relation between the MSD and ACF \eqref{msd2}
and using the solution of the Smoluchowski equation for the
orientations given above in Appendix \eqref{sec:anglePDF}. Here we
review their results shortly.  The discussion is valid both in the
ballistic and in the diffusive regime.

The mean square displacement can be calculated using \eqref{msd2}
which \notiz{for the micro-swimmer with constant speed explicitly reads}
\begin{eqnarray}
\label{eq:fpmeinktk}
\langle \mathbf{r}^2(t)\rangle=v_0^2\int_0^t\int_0^tdt_1dt_2\langle \cos(\phi(t_1)\,-\,\phi(t_2)) \rangle.
\end{eqnarray}
Here $\phi(t_1),\phi(t_2)$ are the orientations of motion at the two
times $t_1$ and $t_2$, respectively, and the average has to be taken
over their probability distribution, Eq.\eqref{orient1}.  Calculating
the integrals leads to the result for the mean square displacement
found by Mikhailov and Meink\"ohn:
\begin{eqnarray}
  \langle \mathbf{r}^2(t)\rangle=\frac{2v_0^4}{D_{\phi}}\,t\,+\,\frac{2\,v_0^6}{D_{\phi}^2}\,\left[\exp\left(\,-\,\frac{D_{\phi} t}{v_0^2} \right)\,-\,1 \right]
\end{eqnarray}
Similarly to Eq.\eqref{msd1}, it describes the ballistic and the
diffusive behavior. For times $t \,\gg \,\tau_{\phi}$ the second term
can be neglected and the MSD scales linearly in time with the
diffusion coefficient given by Eq.\eqref{eq:eff_diff}. Linear growth
in time is obtained at a coarse grained scale $|\mathbf{r}| \gg l_{\phi}=
v_0 \tau_{\phi}= v_0^3/D_{\phi}$.

\end{document}